# Development of a Full Monte Carlo Therapeutic Dose Calculation Toolkit for Halcyon Using Geant4

Ruirui Liu[1], Zhen Ji[2], Xiandong Zhao[2], Tianyu Zhao[2], Abhishek Sethi[3], Daren Sawkey[4], Bin Cai[5]

[1]Department of Radiation Oncology and Winship Cancer Institute, Emory University, Atlanta, Georgia, 30322, USA

[2]Department of Radiation Oncology, Washington University School of Medicine, St. Louis, MO, 63110, USA

[4]University of Illinois-Chicago College of Medicine, Chicago, IL, 60612, USA

[4]Varian Medical Systems, Inc., 3120 Hansen Way, Palo Alto, California 94304

[5]Department of Radiation Oncology, University of Texas Southwestern Medical Center, Dallas, TX 75287, USA

**Abstract:**

**Purpose:** To develop a Monte Carlo (MC) therapeutic dose calculation toolkit of a recently released ring gantry linac in Geant4 (Version 10.7) for secondary dose validation of radiotherapy plan.

**Methods:** Halcyon (Varian Medical Systems), a novel ring gantry linac equipped with double stack multi-leaf collimator (DSMLC) system was recently released. The DSMLC was modeled and radiation transport in DSMLC and patient phantom was simulated using Geant4. Radiation source was sampled from a phase space file for linac head above the DSMLC. The phase sparce file was obtained using a cloud-based Monte Carlo (MC) simulator, VirtuaLinac (VL) provide by Varian. Dosimetric profiles for different square field widths (2x2, 4x4, 6x6, 8x8, 10x10, 20x20, and 28x28 $cm^2$), i.e., percent depth dose (PDD) curves and lateral profiles are simulated and compared against the experimental profiles. IMRT (intensity modulated radiation therapy) plans in two anatomical sites (prostate and brain) were also calculated using the developed toolkit and compared against the TPS calculated dose (Acuros, Eclipse 15.6). 3D dose difference and 3D gamma analysis were used to evaluate the simulation accuracy compared against the TPS calculated dose.

**Results:** The simulated lateral dose profiles and PDD curves in water phantom match well with the measured ones for all the simulated field sizes with relative difference ±2%. For the prostate and brain IMRT plans, the simulated dose showed a good agreement with the TPS calculated dose. The 3D gamma pass rate (3%/3mm) are 98.08% and 95.4% for the two prostate and brain plans, respectively.

**Conclusion:** The developed full MC dose calculation toolkit for Halcyon performs well in dose calculations in water phantom and patient CT phantom. The developed toolkit shows promising possibility for future secondary dose calculation for IMRT and serve as clinical quality assurance (QA) tool for Halcyon.

# 1. Introduction

In 2017, a novel ring gantry linac, Halcyon (Varian Medical Systems, Inc., Palo Alto, CA, USA), was released aiming at enhancing treatment efficiency while providing high quality radiotherapy [1]. The jawless design, rapid gantry rotation and use of double stack multi-leaf collimator (DSMLC) are novel features of the Halcyon, and it is a straight-through delivery that uses only DSMLC to collimate the 6 MV flattening-filter-free beam [2]. Monte Carlo method has been used to conduct radiation transport simulation for linac because it can simulate stochastic radiation interaction process in complicated machine geometry with high order of accuracy [3,4]. Many simulation studies have been done on modeling different linacs [5–9] using a variety of simulation code, such as MCNP [10], EGS4 [6], EGSnrc [11], Geant4 [12] and PENELOPE [13]. However, there still is no Monte Carlo simulation for the newly released Halcyon linac with complex DSMLC. For modeling double stack MLC, a recent work has been done for ViewRay MRI Linac using Geant4 [14], but no patient treatment plan was calculated using their simulation code. The overall goal of this work is to develop a full Monte Carlo dose calculation toolkit for Halcyon with the capability of simulating radiation transport of 6 MV flattening-filter-free (FFF) beam in DSMLC and patient dose calculation with patient CT dataset.

# 2. Methods and Materials

## 2.1 Geant4 simulation

Geant4 is an available software used to perform Monte Carlo simulations of the interactions between energetic particles in matter [15,16]. In the field of medicine simulation, Geant4 has been widely tested and benchmarked against the other popular radiation transport codes such as EGS and MCNP [17]. In this work, Geant4 (Version 10.7) was used to simulate radiation transport in DSMLC calculate patient dose with patient CT phantom. The physics list, EMStandardOpt4 [18], was used for radiation transport simulation. The default range cut value was set as 1 μm.

## 2.2 Beam Model from VirtuaLinac

For accurate modeling of the linac using the Monte Carlo method, we must have precise knowledge of the dimensions and materials of the linac head assembly components that influence the beam. This information is generally provided by the linac vendors, but it is not available for the Halcyon linac. Instead, Varian provides an online Monte Carlo linac simulator based on Geant4, VirtuaLinac (Varian Medical System), to assist users to simulate the radiation transport process inside the linac head. The VirtuaLinac is currently hosted on Amazon Web Service (AWS), an online cloud-computing platform, and can be initiated via user command. The VirtuaLinac has been validated and used in several investigations [19–21]. In this work, we used the VirtuaLinac (Version 1.4.12) to obtain an International Atomic Energy Agency (IAEA)-compliant phase space file just above the DSMLC [22]. In simulation using VirtuaLinac, no variance reduction technique was used and the splitting factor for Bremsstrahlung was set to 1. The phase space file was generated by simulating 6 billion electrons with energy 6 MeV, which was taken about 1 week by using eight standard computation cores in Amazon cloud. The generated phase file was used to sample the radiation source in Geant4 for dose calculation.

## 2.3 Modeling DSMLC

The DSMLC consists of two MLC layers, the distal and proximal layer. The leaves are composed of 95% tungsten alloy and the remaining 5% is not specified by the alloy manufacturer. For our simulation, we assume that the 95% tungsten alloy of DSMLC is the same alloy for the MLC of TrueBeam (Varian) whose material information is provided in the TrueBeam Monte Carlo package version 1.1 available at **www.myvarian.com**. The TrueBeam MLC is made of tungsten alloy (MIL-T-21014, type II, class 3, non-magnetic), and the combination was: 1.5% Cu and 3.5% Ni. The Halcyon MLC leaf is 7.7 cm thick and 1 cm wide (when projected at isocenter) with a leaf end curvature radius of 23.4 cm [1]. There is a 0.5cm offset between distal and proximal layer. The upper layer has 29 leaves and lower layer has 28 leaves with the maximum field size of 28 cm×28 cm. In Halcyon 2.0, both layers can modulate the field independently. Schematic diagrams of the simulated DSMLC are shown in Figure 1.

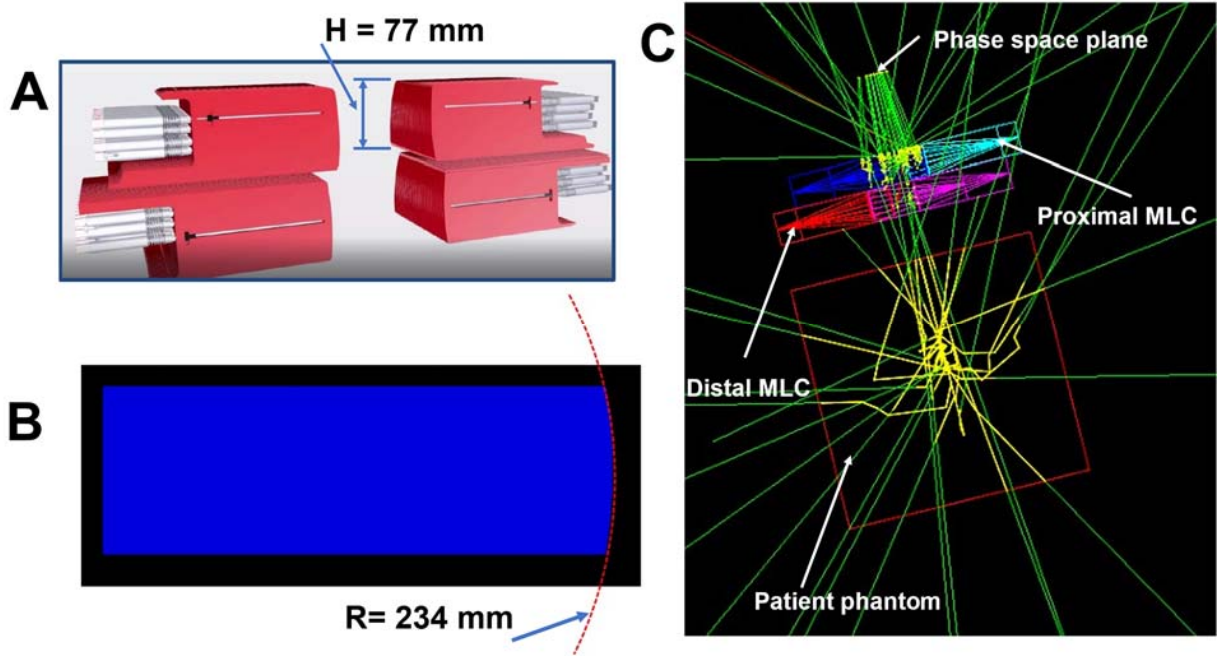

**Figure 1**: Modeling of DSMLC in Geant4. A. Schematic of DSMLC of Halcyon. B. A simulated single MLC leaf rendered in Geant4. C. Schematic of radiation transport simulation for DSMLC and patient phantom using Geant4.

2.4 Open field simulation

For Halcyon, the open fields are defined by DSMLC. In this work, radiation transport for a few field sizes (2×2, 4×4, 6×6, 8×8, 10×10, 20×20, and 28×28 $cm^2$) are simulated. The leaves are static in simulation and the dose in a water tank with dimension as 40×40×40 $cm^3$ is simulated. The dose tally voxel size is set as 4×4×4 $mm^3$. The percent depth dose (PDD) curves at a source-to-surface distance (SSD) of 90 cm and lateral dose profiles of 40×40 $cm^2$ field at 10 cm depth are obtained. $1\times10^8$ histories are simulated for each field size. The simulated profiles are compared with the experimental profiles and relative difference are obtained.

2.5 Dose measurement

Dosimetric data is obtained using a 3D scanning system (IBA, blue phantom 2) with two CC04 chambers (IBA Dosimetry, active volume: 0.04 $cm^3$) used as reference and scanning chambers. Both beam profiles (crossline and inline at 10 cm depth) and PDDs (along central axis) curves are measured with various field size ranging from 2×2 $cm^2$ to 28×28 $cm^2$. In all measurements, the SSD was 90 cm and both gantry and collimator angles were kept at 0 degrees. The scanning speed is set to be 0.4cm/s and the dose rate is 800MU/min.

## 2.6 IMRT simulation

The dose distributions of four clinical plans are recalculated and compared with the dose distribution calculated by treatment planning system (TPS) (Acuros, Eclipse 15.6). Two prostate IMRT plans and two brain IMRT plans (anonymized patients dataset derived from a study previously authorized by IRB) are exported from TPS in DICOM format with CT DICOM files, plan DICOM files, structure DICOM files, and dose DICOM files. In simulation, the developed dose calculation toolkit imports patient CT DICOM files and plan DICOM files to construct the simulation geometry for the treatment. Three-dimensional voxelized phantom is created based on patient's CT files. The mass density assigned to each voxel is estimated by interpolating from its density vs. HU scanner calibration curve. Then ICRU Report 46 compositions are assigned to each voxel according to a published lookup table [23,24] to build the cross-section files.

The recalculated brain plan is composed of 9 beams with a total of 1162 segments, and the recalculated prostate plan is composed of 9 beams with a total of 1494 segments. Total $4\times10^9$ histories are simulated for each plan. The calculation is implemented in a 40-core workstation (Precision 7920 Tower, Dell Inc.) with total CPU time around 24 hours. The dose distribution is calculated based on the voxelized patient phantom with a voxel size of $1.7578\times1.7578\times1.5$ mm$^3$. Doses of individual segments are summed up for each voxel according to the corresponding monitor units (MU). The calculated dose values of all voxels are saved in a Comma Separated Values (CSV) file after simulation. Then, an in-house developed MATLAB (MathWorks, Natick, MA) code is used to read the dose CSV file and dose DICOM files from the TPS, then the dose analysis is implemented through the MATLAB code. The dose obtained by the developed toolkit is compared with the TPS calculated dose using a gamma criterion [25] of 3%/3mm of the dose maximum with a pass criterion of $\gamma < 1$.

## 2.7. Code repository

In order to achieve code reuse, we share the code including the parameter and document in an open-source software repository (https://github.com/forgetsummer/Halcyon).

## 3. Results

### 3.1 Open Field Simulation

Figure 2 shows the simulated and experiment PDD curves and corresponding comparison results between the two for field sizes of 2×2, 6×6, 10×10, and 28×28. The results for field sizes of 4×4, 8×8, and 20×20 is shown in Fig. S-1 in supplementary. Using current simulation parameters, the simulated PDD curves and lateral dose profiles match well with the corresponding experimental data with relative difference within ±2% for all the simulated fields. For PDD curves, noticeable relatively large difference (5% to 6%) appears at the dose points in buildup region. For lateral profiles, relatively large difference (about 5%) appears at the dose points at field edges.

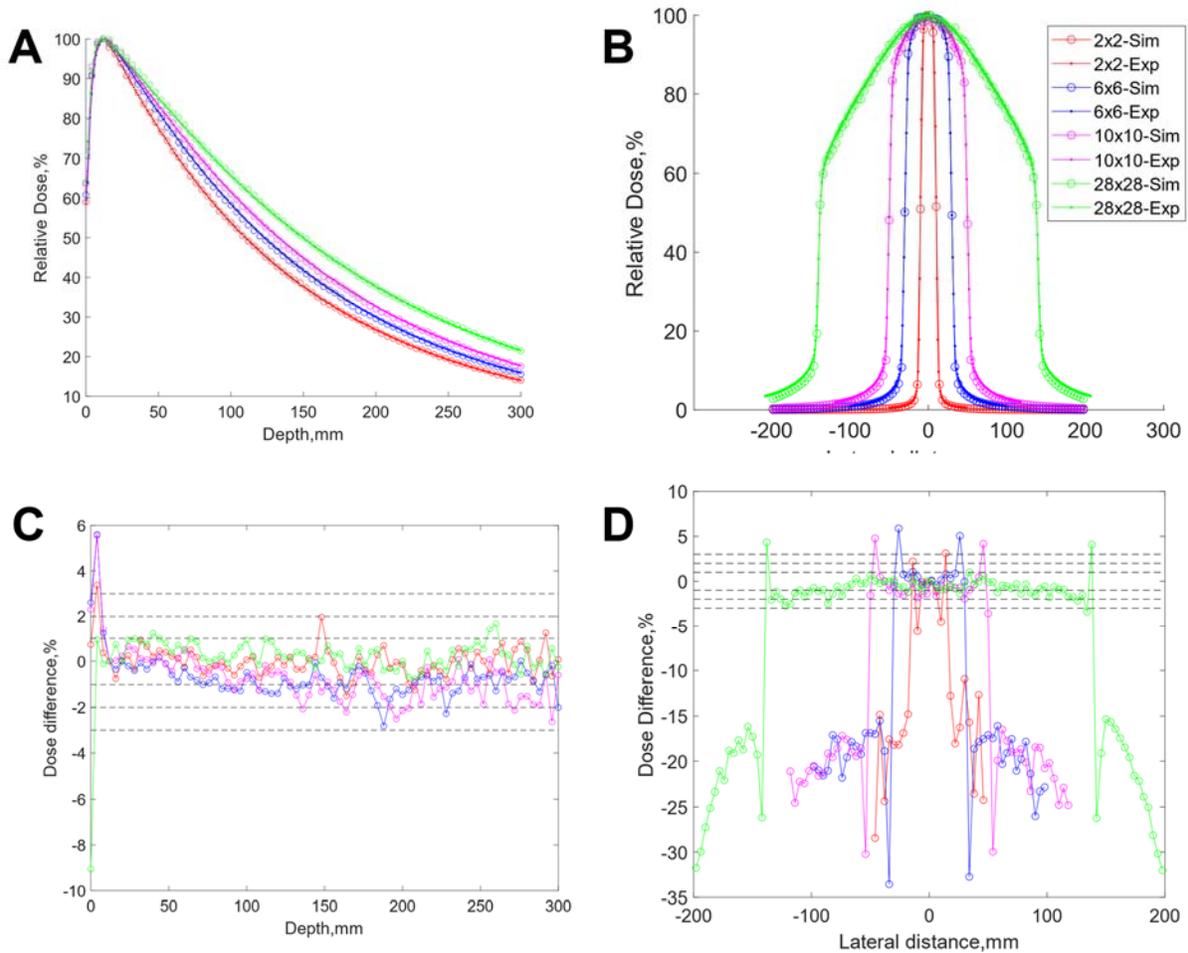

**Figure 2**: Comparison between the simulated and experimental dose profiles. A. Simulated and experimental PDD curves. B. simulated and experimental lateral dose profiles. C. Difference between the simulated and experimental PDD curves. D. Difference between the simulated and experimental lateral profiles.

## 3.2 IMRT Simulation

Figure 3 shows the calculated dose distribution of a prostate IMRT plan. The slice dose in axial, sagittal, and coronal planes of the calculated plan and TPS plan are plotted parallelly in two columns respectively for comparison. The dose difference distribution and gamma index of the two plans are shown in Figure 4. The three-dimensional gamma pass rate for this plan is 98.5%. The 3D dose difference is marginally small in whole treatment area except body surface at the beam entrance region where the dose difference is roughly about 20% to 30%. The gamma index failing voxels appear in high dose gradient region. The calculated dose distribution information for another prostate IMRT plan has a gamma pass rate of 95.1%, and the dose distribution and analysis are shown in Fig. S-2 and Fig. S-3 respectively in supplementary. The dose distribution pattern is similar as the first IMRT plan.

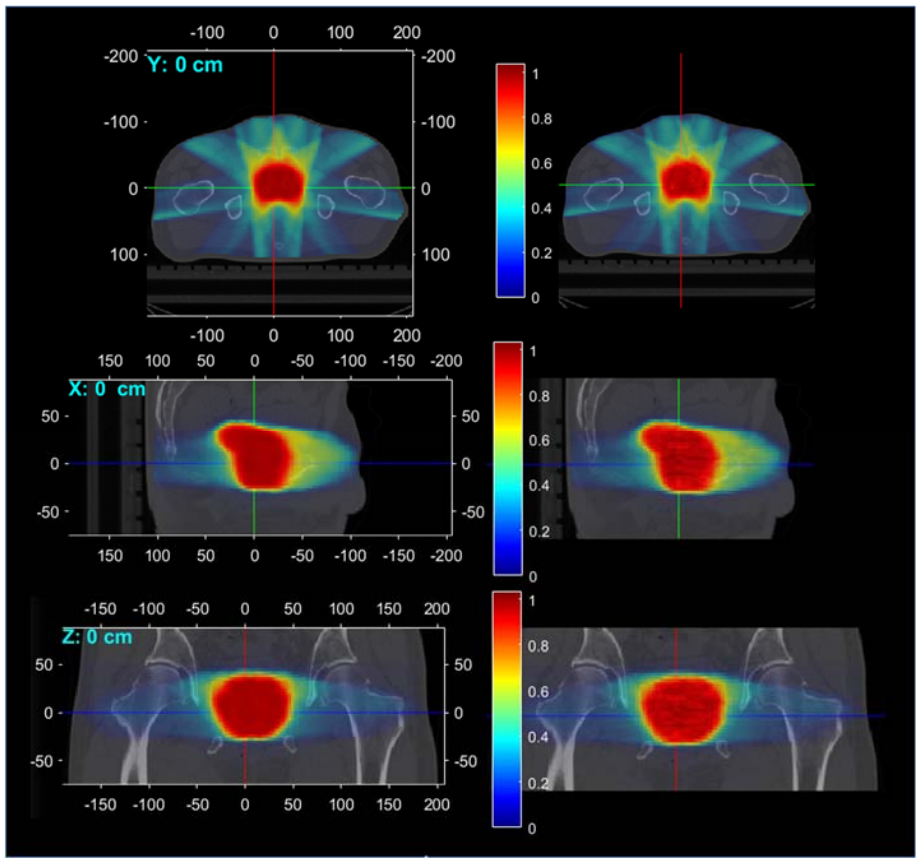

**Figure 3**: Calculated dose distribution of a prostate IMRT plan. The first column is TPS calculated dose distribution in axial, sagittal, and coronal plane, respectively. The second column is MC calculated dose distribution in axial, sagittal, and coronal plane, respectively.

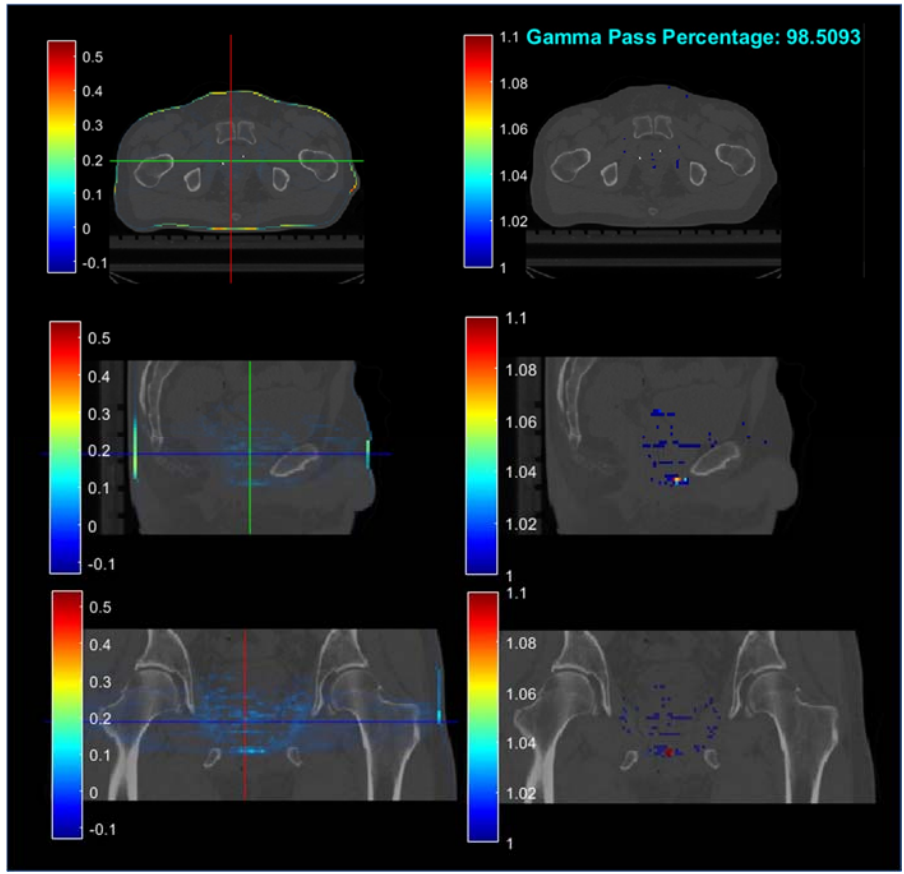

**Figure 4:** Calculated dose analysis of the prostate plan in **Figure 3**. The first column is dose difference between dose of TPS and MC calculated dose. The difference is quantified as relative difference. The second column is gamma index distribution in axial, sagittal, and coronal plane, respectively.

Figure 5 shows the calculated dose distribution of a brain IMRT plan, and Figure 6 shows the corresponding dose difference and gamma index distribution. The three-dimensional gamma pass rate for this plan is 96.6%. Another brain IMRT plan has a gamma pass rate of 93.3%, and calculated dose distribution information are shown in Fig. S-4 and Fig. S-5 respectively in the supplementary.

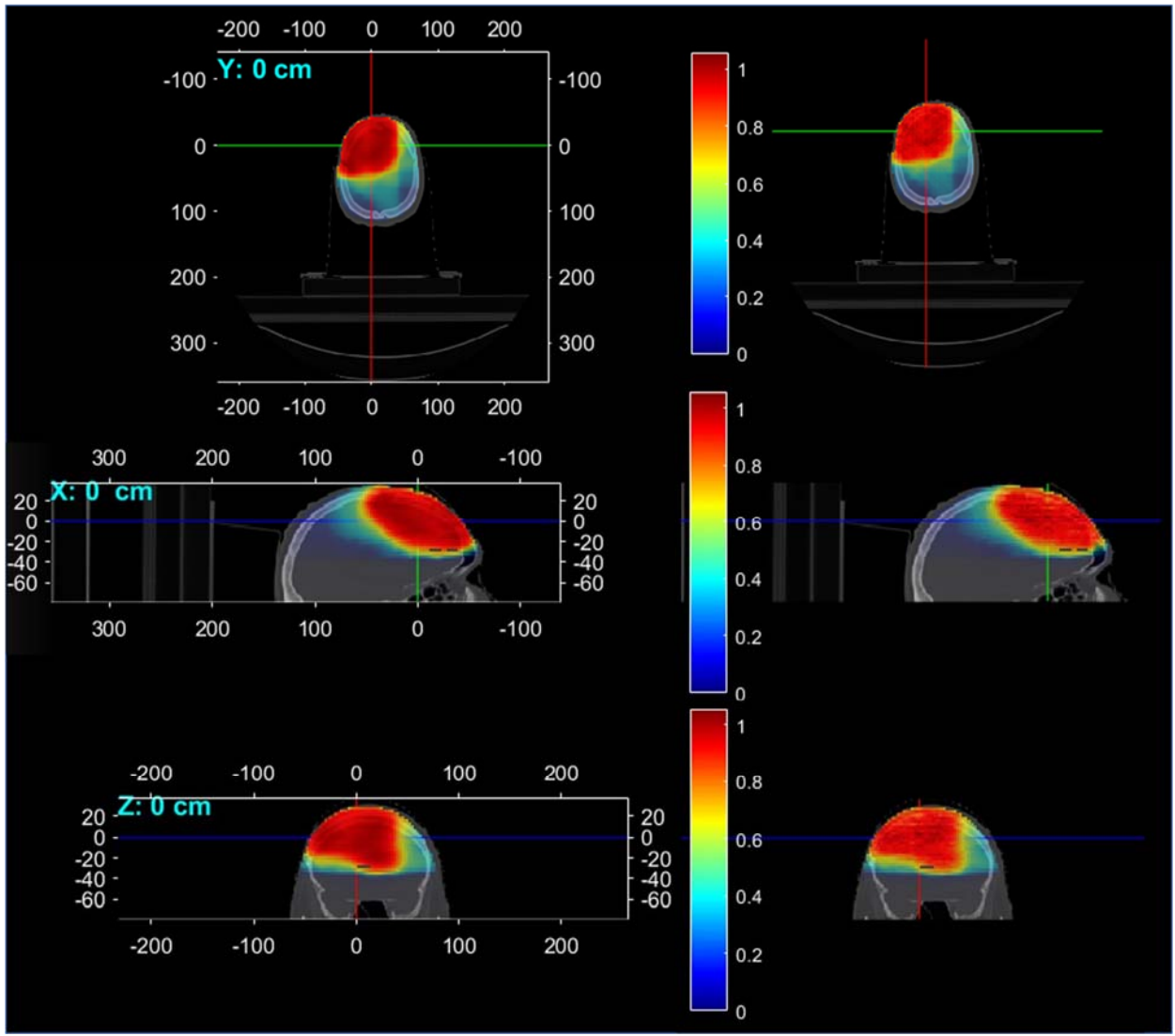

**Figure 5**: Calculated dose distribution of a brain IMRT plan. The first column is TPS calculated dose distribution in axial, sagittal, and coronal plane, respectively. The second column is MC calculated dose distribution in axial, sagittal, and coronal plane, respectively.

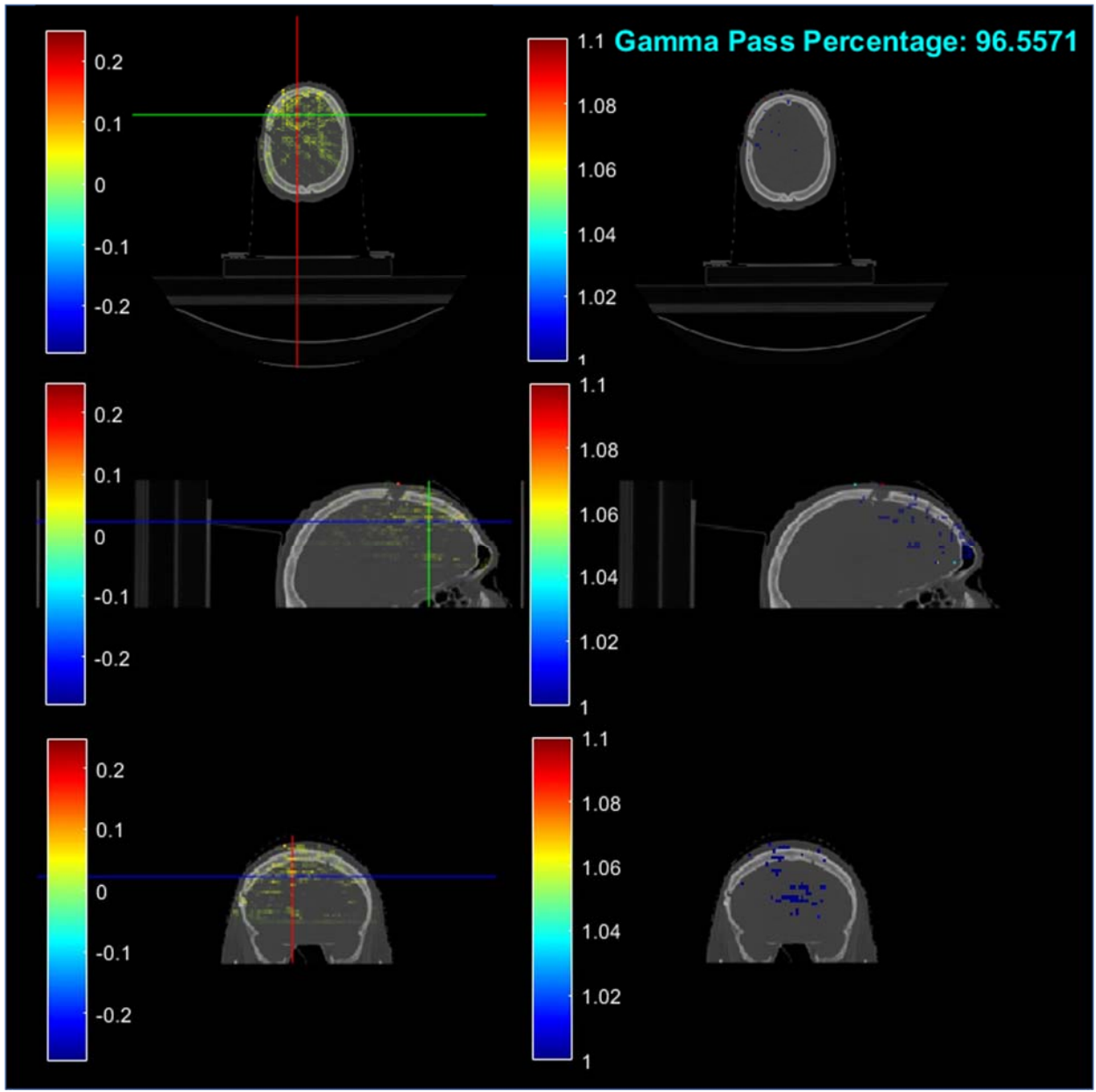

**Figure 6**: Calculated dose analysis of the brain plan in **Figure 5**. The first column is dose difference between dose of TPS and MC calculated dose. The second column is gamma index distribution in axial, sagittal, and coronal plane, respectively.

## 4. Discussion

In this study, a full Monte Carlo dose calculation toolkit for Halcyon was developed in Geant4. The developed toolkit is capable of computing absorbed dose distributions in CT phantom of patients treated with the Halcyon linac. We used online tools (VirtuaLinac) distributed by the manufacturer to its clients for tallying a phase-space file upstream of the collimating devices of the linac. This phase-space file was subsequently used in the code developed to simulate radiation transport either in a water phantom or in the CT phantom of patients. To our best knowledge, this is the first work done on the computation of dose distributions in patients irradiated with this linac model.

Currently, Varian does not disclose information on the geometrical description of the linac head of Halcyon. The information used in this work for modeling the collimators is approximate and does not reflect all the geometrical details of these devices. VirtuaLinac compensates for the inconvenience of lacking accurate detailed information from vendor for modeling beam head of Halcyon linac. We tried to develop a full Monte Carlo model considering all the components of Halcyon with our developed DSMLC model and CT phantom radiation transport model in Geant4 along with usage of VirtuaLinac. By using this design philosophy, we can make sure the developed Monte Carlo dose calculation toolkit reflects the most accurate information this linac given the current state of very limited information available to researchers on the description of this linac.

The AAPM TG-105 report [26] classifies the approaches of Monte Carlo modeling MLC into three categories, i.e., pseudo-explicit, explicit transport, and explicit approximate transport [26]. Our developed model for DSMLC belongs to explicit transport category, as it can explicitly account for leakage radiation and MLC scatter radiation during the dose calculation simulation [27]. One limitation of explicit transport model is its high computational cost. In future studies, we can attempt to enhance the computational efficiency, such as by not accounting for the detailed transport process of scattered Compton photons and electrons in MLC [28].

For most of the calculated PDD curves and lateral dose profiles in water tank, the difference between the simulated and experimental measurements was within ±2%. The edge positions of the simulated lateral dose profiles all matched well with the experimental profile, which indicated that the DSMLC model works well for simulating the radiation transport of photon in MLC configuration for shaping different field sizes.

Only IMRT plans are currently included in this preliminary study, and the comparison of VMAT cases will be included in our future work. The calculated dose distribution of two IMRT plans of prostate tumor cases have high overall gamma pass rate (>95%). One possible reason for the failing gamma index in some voxels is due to the steep dose gradients in high dose region. This is similar to Friedel's work for calculating the dose distribution prostate IMRT plan using Monte Carlo method [29]. Another possible reason is due to variation of electron density used in TPS and the developed dose calculation toolkit.

One possible reason for the dose difference in the region of patient's skin is due to the differences in treatment of air tissue surface in TPS and in Monte Carlo dose calculation. The lack of precision of the TPS at the air-tissues interface and the strong dose gradient of the build-up region can be challenging to compute surface dose [30]. For example, the AXB algorithm can lead to up 3.6% difference compared to the Monte Carlo method in calculating the dose in air/tissue region [31]. As expected, higher accuracy is seen for the relatively homogeneous prostate, whereas more discrepancies are observed for the more heterogeneous head case (where tumor close to the skull). Some brain tumors have more heterogeneous tissue areas (i.e., bone and air cavities) than prostate tumors, complicating accurate modeling of radiation dose deposition.

The computation time of this toolkit for IMRT plans is compatible with the main objective of this toolkit, that it is not to be used universally in patient care but only for complex cases. In future study, we can improve computational efficiency by adapting additional variance reduction techniques ranging from very simple, e.g., survival biasing and forced collision, to more complex, e.g., randomly allocated next-flight estimation [32].

## 5. Conclusions

A full Monte Carlo dose calculation toolkit has been developed to conduct patient dose calculation for Halcyon linac using Geant4. The preliminary results demonstrated that the developed toolkit is capable of obtaining dose distribution of clinical IMRT plan, and the calculated dose is in strong agreement with the TPS calculated dose. The developed toolkit shows promising possibility for future secondary dose calculation for IMRT and serve as clinical QA tool for Halcyon.

# Supplementary

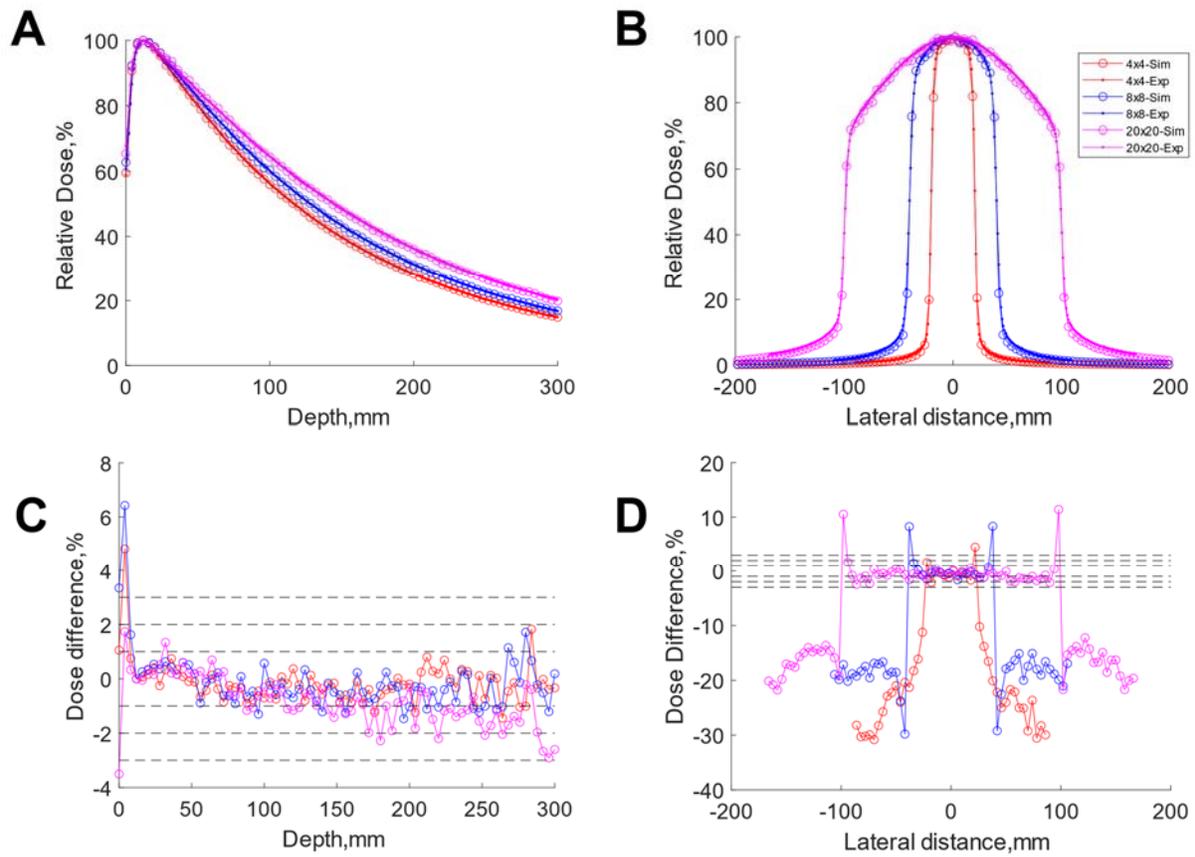

**Figure S-1**: Comparison between the simulated and experimental dose profiles. A. Simulated and experimental PDD curves. B. simulated and experimental lateral dose profiles. C. Difference between the simulated and experimental PDD curves. D. Difference between the simulated and experimental lateral profiles.

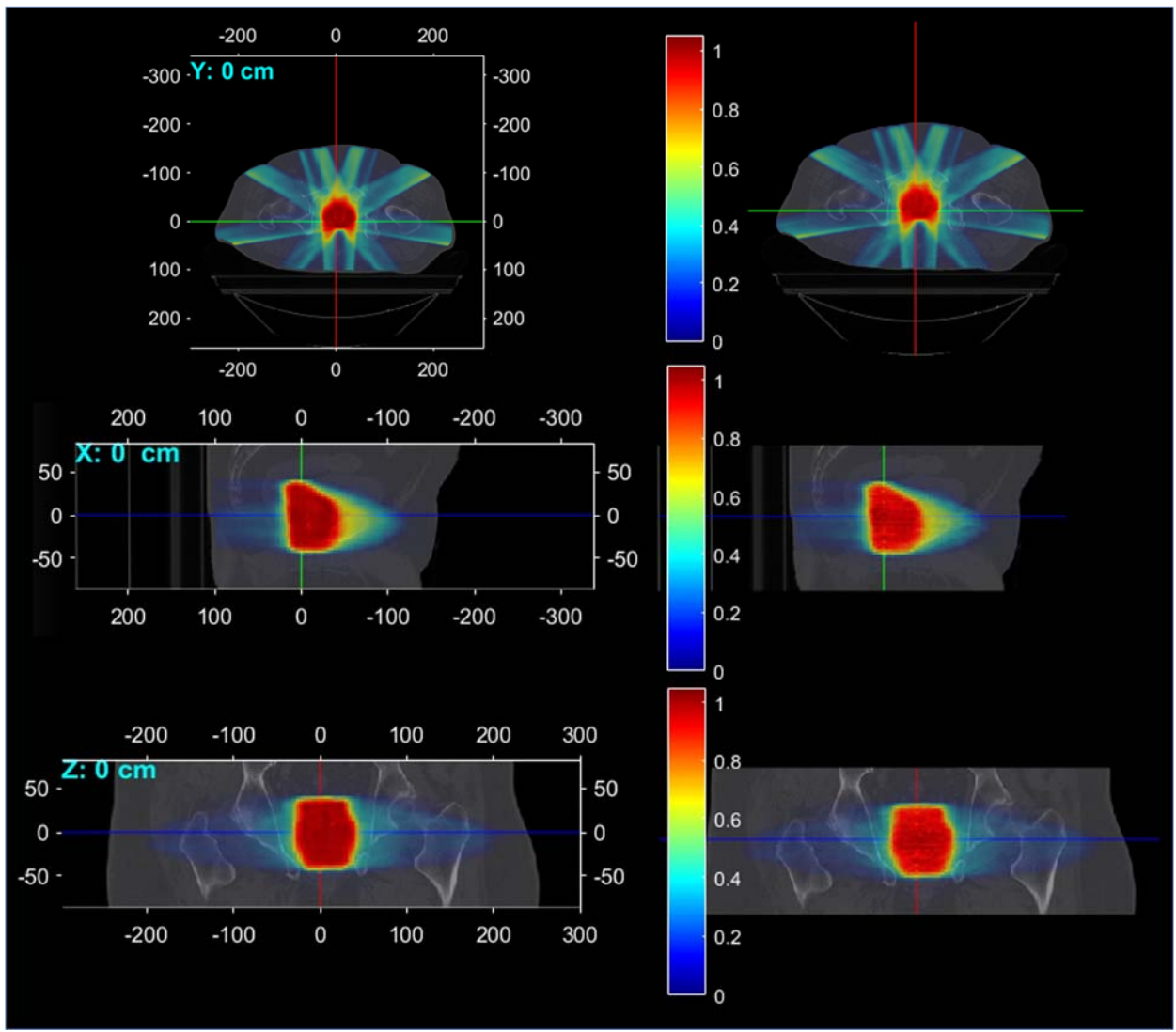

**Figure S-2:** Calculated dose distribution of a prostate IMRT plan. The first column is TPS calculated dose distribution in axial, sagittal, and coronal plane, respectively. The second column is MC calculated dose distribution in axial, sagittal, and coronal plane, respectively.

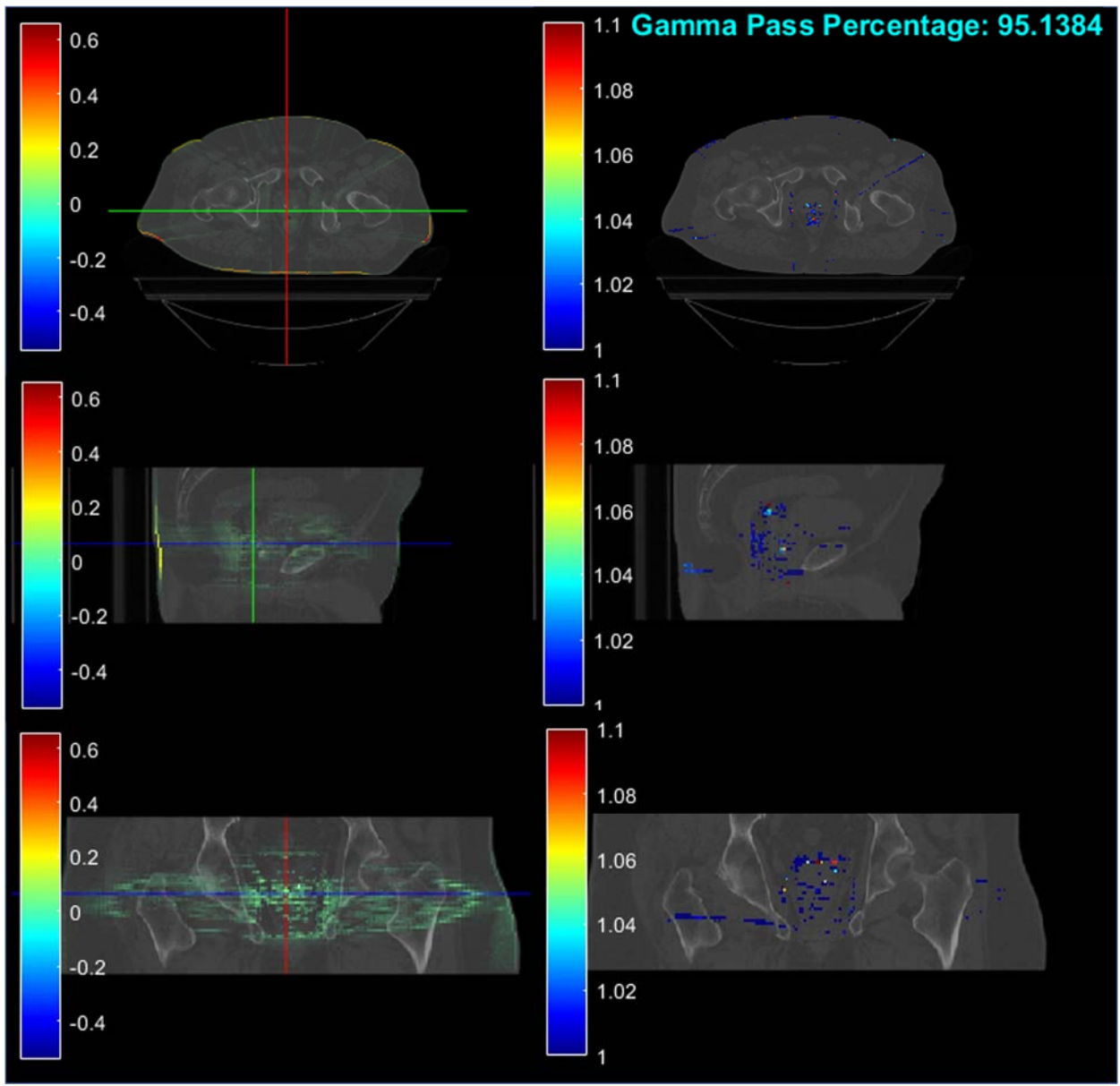

**Figure S3**: Calculated dose analysis of the prostate plan in Figure S2. The first column is dose difference between dose of TPS and MC calculated dose. The second column is gamma index distribution in axial, sagittal, and coronal plane, respectively.

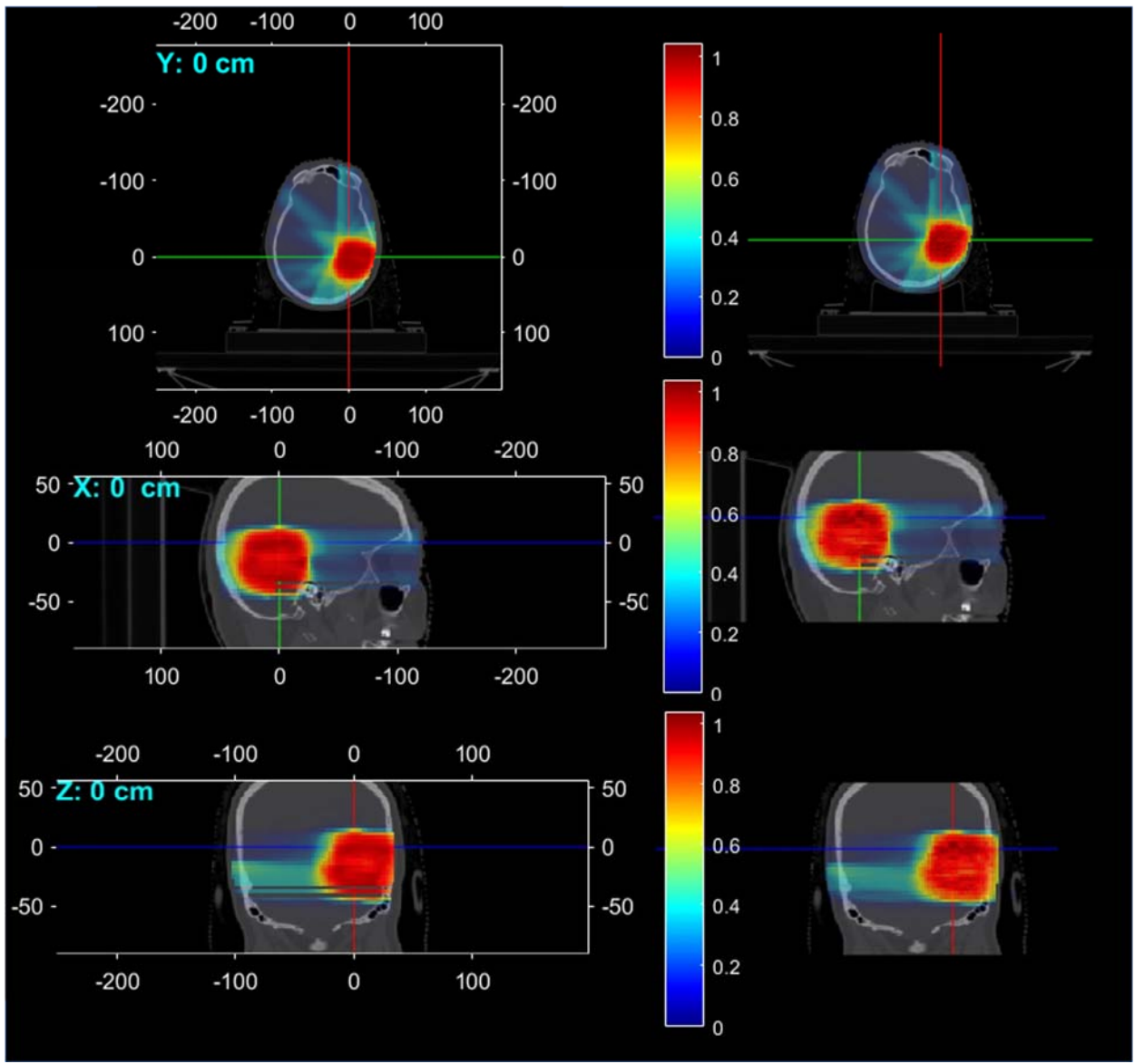

**Figure S-4**: Calculated dose distribution of a brain IMRT plan. The first column is TPS calculated dose distribution in axial, sagittal, and coronal plane, respectively. The second column is MC calculated dose distribution in axial, sagittal, and coronal plane, respectively.

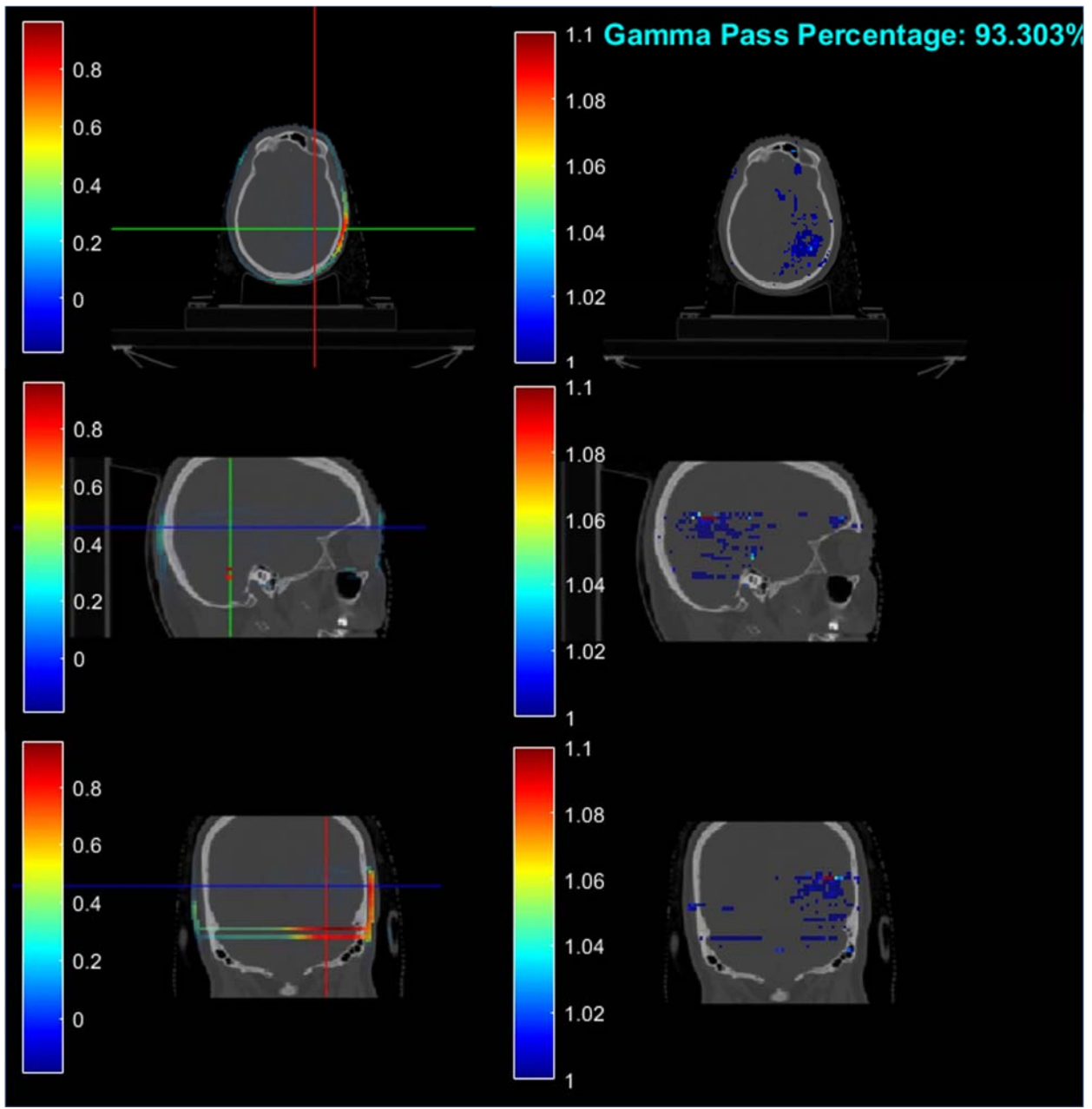

**Figure S-5**: Calculated dose analysis of the brain plan in Figure S4. The first column is dose difference between dose of TPS and MC calculated dose. The second column is gamma index distribution in axial, sagittal, and coronal plane, respectively.